# Electric-field control of exchange bias in multiferroic epitaxial heterostructures

V. Laukhin[1,2], V. Skumryev[2,3], X. Martí[1], D. Hrabovsky[1], F. Sánchez[1], M.V. García-Cuenca[4], C. Ferrater[4], M. Varela[4], U. Lüders[5], J.F. Bobo[5], and J. Fontcuberta[1]

[1] Institut de Ciència de Materials de Barcelona-CSIC, Campus UAB, Bellaterra 08193, Spain

[2] Institut Català de Recerca i Estudis Avançats (ICREA), Barcelona, Spain

[3] Departament de Física, Universitat Autònoma de Barcelona, Bellaterra 08193, Spain

[4] Departament de Física Aplicada i Òptica, Universitat de Barcelona, Diagonal 647, Barcelona 08028, Spain

[5] LNMH ONERA-CNRS, BP 4025, 31055 Toulouse, Cedex 4, France

ABSTRACT

The magnetic exchange bias between epitaxial thin films of the multiferroic (antiferromagnetic and ferroelectric) hexagonal $YMnO_3$ oxide and a soft ferromagnetic (FM) layer is used to couple the magnetic response of the ferromagnetic layer to the magnetic state of the antiferromagnetic one. We will show that biasing the ferroelectric $YMnO_3$ layer by an appropriate electric field allows modifying and controlling the magnetic exchange bias and subsequently the magnetotransport properties of the FM layer. This finding may contribute to pave the way towards a new generation of electric-field controlled spintronics devices.

Multiferroic materials have been proposed to allow building a new generation of devices in spintronics, eventually allowing overcoming critical limitations in technology [1,2]. Much of effort has been directed to search for materials displaying the elusive coexistence of ferroelectric (FE) and ferromagnetic (FM) behavior [3,4], which is thought to be essential for progress in this direction. On contrast, materials displaying coupled FE and antiferromagnetic (AF) behavior have received much less attention. At first sight this may appear to be surprising as AF materials are nowadays used in many magnetic devices. Essential for optimal exploitation of the multiferroic character of a material is that the ferroic properties (magnetic and electric, in the present context) are coupled. Hexagonal $YMnO_3$, in bulk form, is ferroelectric up to 900 K and exhibits an antiferromagnetic character at low temperature ($T_N \sim 90$ K). It has been recently shown that in $YMnO_3$ single crystals both order parameters are coupled [5] and this observation has triggered a renewed attention to this oxide as candidate in multifunctional heterostructures [6,7]. The electric polarization axis of $YMnO_3$ is along the c-axis of the hexagonal structure; the Mn atomic spins lie in a perpendicular plane, forming a two dimensional, frustrated antiferromagnetic, triangular network [8, 9].

It is thus clear that, in principle, it could be possible to use AF $YMnO_3$ to pin the magnetic state of a suitable magnetic material and subsequently exploit its ferroelectric character and the coupling between FE and AF order parameters to tailor the properties of the FM layer. As a first step, it has been recently shown that indeed it is possible to exchange-bias NiFe (Permalloy-Py) with antiferromagnetic epitaxial (0001) $YMnO_3$ thin films which displays a remanent electric polarization [6].

Attempts towards electric field control of exchange bias have been recently reported by Borisov et al. using magnetoelectric, but not multiferroic, (AF) $Cr_2O_3$ single crystals as pinning layers [10]. Indeed, it has been shown that using an appropriate electric field cooling process the antiferromagnetic domains of $Cr_2O_3$ can be modified with subsequent changes in the exchange bias.

Here we will show that by using appropriate materials, it is possible to grow heterostructures that, exploiting the AF character of $YMnO_3$, allows control of the magnetic state of a ferromagnetic layer by an electric field. For that purpose, thin epitaxial layers of $YMnO_3$ have been sandwiched between metallic electrodes (Pt and Py) and the exchange bias between $YMnO_3$ and Py has been monitored as a function of a biasing electric field applied across the $YMnO_3$ layer (Fig. 1b).

Exchange bias at interface between ferromagnetic and antiferromagnetic materials is recognized to be associated to the existence of a net magnetization at the surface of the AF and to the existence of a unidirectional magnetic anisotropy that pins the magnetization of an upper-grown ferromagnetic layer [11]. As a consequence, when a magnetic field is applied parallel to the interface, the magnetization of the ferromagnetic layer does not follows (neglecting the anisotropy of the FM layer) the external field $\mathbf{H_a}$ but the $\mathbf{H_a}+\mathbf{H_{eb}}$ vector sum, where $\mathbf{H_{eb}}$ is the exchange bias field. This behavior, and thus the presence of a finite $H_{eb}$, is most commonly evidenced by a shift along the magnetic field axis of the magnetization loop of the FM layer. However, it also dramatically affects other properties, such as the angular dependence of anisotropic magnetoresistance (AMR) of the FM layer when the external magnetic field is rotated. This technique has been already used to measure the exchange bias in FM/AF bilayers [6, 12]. If $\theta_M$ is the angle between the magnetization (**M**) and the measuring electric current direction (**J**), the resistivity is given by $\rho(\theta_M) = \rho_\perp + \Delta\rho\cos^2\theta_M$, where $\Delta\rho \equiv \rho_{//} - \rho_\perp$ and $\rho_\perp$ and $\rho_{//}$ are the resistivity for $\mathbf{M}\perp\mathbf{J}$ and $\mathbf{M}//\mathbf{J}$, respectively. In presence of $H_{eb}$, when rotating the external field $\mathbf{H_a}$ to an angle $\theta_a$ with respect to **J**, the measured $\rho(\theta_a)$, in general, does not follow a simple quadratic $\cos^2(\theta_a)$ dependence. Thus departure from this simple behavior allows monitoring the existence of $H_{eb}$ and its eventual modifications under a biasing electric field.

$YMnO_3$ (0001) films, 90 nm thick, with hexagonal structure have been grown by pulsed laser deposition on $SrTiO_3$ (111) substrates buffered with a thin epitaxial Pt layer (8nm) as bottom metallic electrode. This heterostructure was covered by a Py film (15nm). X-ray diffraction experiments indicate that the Pt and $YMnO_3$ films were fully c-axis textured whereas in plane, films were fully epitaxial with out-of-plane cell parameters $d_{111}(Pt) = 2.27$ Å and $c(YMnO_3) = 11.45$ Å. Extensive structural details are reported elsewhere [13]. During the growth of $YMnO_3$, a mask was used -partially covering the bottom Pt electrode- for subsequent electric contacting.

Four (in-line) electric contacts on Py were used for transport measurements. Additional electrical contacts on Py and Pt were made for electric biasing the $Py/YMnO_3/Pt$ sandwich. The room-temperature resistivity of the $YMnO_3$ layer is of about $10^6$ $\Omega$cm.

Anisotropic magnetoresistance is measured by rotating an external field $\mathbf{H_a}$ in the plane of the sample by using a PPMS from Quantum Design. A sample holder of a SQUID magnetometer has been adapted to allow measurement of the sample magnetization while biasing it by an electric field.

In Fig. 1a we show the magnetization loops, measured at 2 K, under various bias-voltages ($V_e$), after field-cooling (3 kOe) the sample from room-temperature. We notice, in the unbiased sample ($V_e = 0$), the existence of a clear shift of the loop which reflects the existence of an exchange bias field $H_{eb}$ (≈60 Oe). This observation is compatible with the AF nature of the $YMnO_3$ thin film below $T_N$ (~90 K). In order to monitor the temperature dependence of the exchange bias field, we have measured the temperature dependence of the magnetization at $H_a = -100$ Oe, after 3 kOe field cooling. The sample has been heated from 2 K up to 25 K. The results shown in Fig. 1 (inset, top panel) reflect a rapid fall of the magnetization, and thus of the $H_{eb}$ field, upon heating. The rapid fall of $H_{eb}$ with increasing temperature, results from the low in-plane magnetic anisotropy of the hexagonal $YMnO_3$ [6,14]. Notice in Fig. 1a (inset, bottom panel) that, as expected, upon subsequent cooling from 25 K to 2 K the magnetization magnitude ($|M|$) increases.

Data in Fig.1a, clearly show that when applying a bias-voltage across the $YMnO_3$ layer, the shift of the M(H) loop gradually disappear and for $V_e = 1.2$ V the loop appear to be symmetric thus indicating the suppression of $H_{eb}$. Inspection of the M(H) loops in Fig. 1a suggests that it should be possible to reverse the magnetization of the Py upon suitable electric field biasing. A dramatic modification of the magnetic response has been indeed measured (Fig. 1b), at 2K, after cooling the sample under 3 kOe field and fixing the measuring field $H_a = -100$ Oe. The arrow in Fig. 1a illustrates schematically the process. When increasing $V_e$, the magnetization lowers, switches its sign at $V_e ≈ 0.4$ V and virtually saturates at about $V_e ≈ 1.2$ V. Reduction and change of polarity of the electric field does not allow to switch back to magnetization initial M > 0 value, but it remains M < 0. This trend can be well appreciated in the zoom shown in Fig. 1b (left inset); it is remarkable that upon subsequent reduction of $V_e$, the magnitude of M lowers down (about -5%), to a minimum value at $V_e ≈ 0$ and increases again when changing the polarity of the electric field. We notice that this behavior is just opposite to what should be expected if thermal effects associated to Joule heating, virtually unavoidable, were the unique reason for the observed magnetization variation; as illustrated by data in Fig. 1a (inset, bottom panel), the magnetization magnitude should keep growing (about +0.5%) when reducing Joule power ($V_e$). Therefore these measurements indicate that the magnetic exchange bias and thus the system magnetization can be strongly modified by an electric field.

Further evidence will be gained from magnetotransport data. AMR measurements of Py were performed after field cooling (3 kOe) the sample from room-temperature with the magnetic field applied at 45º of the measuring current direction. In this configuration, the AMR is most sensitive to the presence of $H_{eb}$ [12]. Once at the targeted temperature, the magnetic field is reduced down to the measuring field $H_a$ value, and it is afterwards rotated clockwise from -5º to 365º and anti-clockwise to the original position ($\theta_a = 0$º corresponds to **J**//**$H_a$**) while monitoring its resistance $R(\theta_a)$ as a function of the rotating angle ($\theta_a$).

Fig. 2 collects the $R(\theta_a)$ curves, obtained using $H_a$ = 40 Oe, at some selected temperatures (2, 6, 10, 15, 50 and 100 K). Arrows indicate the sense of rotation. We first notice that at 100 K, $R(\theta_a)$ is well described by a $\cos^2\theta_a$ dependence. This implies that at this temperature $\theta_a = \theta_M$ and thus **M** is parallel to **$H_a$**. Indeed, this is what should be expected as $H_{eb}$ is zero in the paramagnetic phase of YMnO$_3$. However, when reducing the temperature, the $\cos^2\theta_a$ dependence disappears illustrating the presence of $H_{eb}$. Only one minimum in $R(\theta_a)$, occurring at about $\theta_a \sim 120$º, remains visible at the lowest temperature. This indicates that at the lowest temperature, under a rotating $H_a$ = 40 Oe field, the film magnetization remains pinned along the $H_{eb}$ direction [6].

We notice that the measurements at 50 K still reflect a clear departure from the $\cos^2\theta_a$ behavior thus indicating that at this temperature, $H_{eb}$ is finite and with a strength comparable to the measuring field ($H_a$ = 40 Oe). This observation contrasts with that extracted from magnetization measurements (Fig. 1a (inset, top panel)) where it was clear that the exchange bias is not longer visible in the magnetization loops above ≥ 10 K. Detailed discussion on the origin of this difference is beyond the scope of this manuscript [6]; we only mention here that it is related to the fact that in magnetization measurements exchange-bias monitoring requires magnetization switching whereas in AMR measurements this is not required. The same reason accounts for the absence of hysteresis in measurements performed at the lowest temperature as the magnetic anisotropy is large and **M** remains confined within an energy minimum while **$H_a$** rotates; at intermediate temperatures hysteresis is evident as **M** is dephased with respect to **$H_a$**. Of course, when approaching the Néel temperature, the anisotropy and thus the hysteresis disappear again. In summary, data of Fig. 2 clearly indicate the existence of a finite $H_{eb}$ field, at least up to 50 K, thus confirming the AF nature of the YMnO$_3$ film.

In order to explore electric field effects on $H_{eb}$ we have measured the AMR response when biasing the Py/YMnO$_3$/Pt sandwich by an electric field. In Fig. 3 we show the data recorded at 5 K (after 3 kOe field-cooling from room temperature) at some selected bias voltages ($V_e$ = 0, 1.2 and 1.8 V). Data in Fig. 3 reveals a strikingly important modification of R($\theta_a$) and thus of exchange bias by $V_e$. We note for instance, that when increasing $V_e$, a second R($\theta_a$) minimum at about ~270º emerges. Comparison of the R($\theta_a$) curves in Fig. 2 and Fig. 3, indicates that upon biasing the $H_{eb}$ becomes relatively less relevant on the R($\theta_a$) response. That is, the electric field effect mimics the effect of increasing the temperature (or alternatively, the applied magnetic field $H_a$ [6]). R($\theta_a$) curves have been recorded for both positive and negative bias-voltages, $V_e$. To provide an overall view of the biasing effect on R($\theta_a$), we collect in Fig. 4 an overall view of the bias-dependence of the R($\theta_a$,$V_e$). For clarity, only clockwise R($\theta_a$) curves have been included and all curves have been shifted fixing, arbitrarily, the deepest R($\theta_a$) minimum to zero. This R($\theta_a$,$V_e$) map gives a clear picture of the suppression of the exchange bias with $V_e$ and the concomitant emergence of a new resistance minimum at $\theta_a \approx$ 270º. Notice that there is a negligible dependence of R($\theta_a$) on the $V_e$ polarity and the electric bias effect is much pronounced at $\theta_a \approx$ 270º. The physical origin of emergence of a minimum of R($\theta_a$) at about $\theta_a \approx$ 270º is illustrated by the diagrams in Fig. 5. Field-cooling with field at an angle of 45º with **J**, determines the uniaxial anisotropy axis defining by the exchange bias field. Thus, for small enough $H_a$ values and for $V_e$ = 0, upon **H$_a$** rotation, **M** remains pinned around the anisotropy axis, thus **M** should be at about $\theta_M \approx$ 45º from **(Fig. 5a, and 5c)**. The appearance of two minima in R($\theta_a$), around $\theta_M \approx$ 90º and 270º, when $V_e$ increases, implies that the **M** gets closer to **H$_a$** thus recovering the cos$^2\theta_a$ dependence **(Fig. 5b and 5d)**. The R($\theta_a$) curves in Figs. 5 have been calculated by using a simple geometrical construction to evaluate the angle between **M** and **J** when rotating **H$_a$** for a fixed **H$_{eb}$**, and subsequently R($\theta_a$). In the simplest approximation, this gives [12]:

$$\rho(\theta) = \rho_{90°} + \Delta\rho \cdot \cos^2\left(\frac{\left(\cos(\theta_{eb}) + \left(\frac{H_a}{H_{eb}}\right) \cdot \cos(\theta_a)\right)^2}{\left(1 + \left(\frac{H_a}{H_{eb}}\right)^2 + 2 \cdot \left(\frac{H_a}{H_{eb}}\right) \cdot \cos(\theta_a - \theta_{eb})\right)}\right) \quad (1)$$

In Figs. 5b and c, $R(\theta_a)$ has been calculated for $H_a/H_{eb}$ = 0.1 and 2 respectively to illustrate the expected modification of $R(\theta_a)$ when reducing the exchange bias. The emergence of two minima in $R(\theta_a)$ upon reduction of $H_{eb}$ is well evident. Therefore, the applied bias voltage has reduced the uniaxial exchange-bias based energy barrier.

The results reported above clearly reveal a genuine electric field effect on the exchange bias in YMnO$_3$/Py heterostructures. The suppression of magnetic exchange bias by electric poling of the underlying YMnO$_3$ ferroelectric layer, indicates a substantial modification of the antiferromagnetic domain structure which is driven by the electric field. The microscopic origin of this effect can not be conclusively inferred from the present experiments; however, although the clamping of ferroelectric and antiferromagnetic domain as reported by Fiebig et al. [5] provides a natural framework to understand the observed effect, more research is required to describe some features such as the observed *reduction* of exchange bias upon electric field biasing and its weak dependence on polarity. We further suggest that similar experiments to those reported here, but performed using biferroic (antiferromagnetic and ferroelectric) oxides having a larger magnetic anisotropy, would allow further improvement of the response and higher operation temperature.

In summary, we have shown that an electric field can be used to tune the exchange-bias coupling in AF/FM heterostructures and eventually the magnetic switching of the FM layer. Here we have used the biferroic antiferromagnetic and ferroelectric YMnO$_3$ layer to bias the FM layer, with the ultimate goal of fully exploiting its ferroelectric character and subsequent hysteretic behavior. Even though, the present results illustrate the promising perspectives of AF/FE oxides in spintronics. These novel heterostructures may have impact on a new generation of electric-field controlled magnetic devices.

Acknowledgments. Financial support by the MEC of the Spanish Government (projects NAN2004-9094-C03 and MAT2005-5656-C04 and FEDER) and by the European Community (project MACOMUFI (FP6-03321)) are acknowledged.

**References**

1. Ch. Binek and B. Doudin, J. Phys.: Conds. Matter **17**, L39 (2005)
2. M. Fiebig, J. Phys. D: Appl. Phys. **38**, R123 (2005)
3. N.A. Hill, J. Phys. Chem. B **104**, 6694 (2000)
4. M. Gajek, M. Bibes, A. Barthélémy, K. Bouzehouane, S. Fusil, M. Varela, J. Fontcuberta, and A. Fert, Phys. Rev. B **72**, 020406(R) (2005)
5. M. Fiebig, T. Lottermoser, D. Frohlich, A.V. Goltsev, and R.V. Pisarev, Nature **419**, 818 (2002)
6. X. Martí, F. Sánchez, D. Hrabovsky, L. Fàbrega, A. Ruyter, J. Fontcuberta, V. Laukhin, V. Skumryev, M.V. García-Cuenca, C. Ferrater, M. Varela, A. Vilà, U. Lüders, and J.F. Bobo , Appl. Phys. Lett. **89**, (2006), in press.
7. J. Dho and M. G. Blamire, Appl. Phys. Lett. **87**, 252504 (2005)
8. see for instance, B.B. van Aken, T.T.M. Palstra, A. Filippetti, and N.A. Spaldin, Nature Mat. **3**, 164 (2004)
9. A. Muñoz, J.A. Alonso, M.J. Martínez-Lope, M.T. Casáis, J.L. Martínez, and M.T. Fernández-Díaz, Phys. Rev. B **62**, 9498 (2000)
10. P. Borisov, A. Hochstrat, Xi Chen, W. Kleemann, and Ch. Binek, Phys. Rev. Lett. **94**, 1176203 (2205)
11. J. Nogués and I.K. Schuller, J. Magn. Magn. Mat. **192**, 203 (1999)
12. B.H. Miller and E. Dan Dahlberg, Appl. Phys. Lett. **69**, 3932 (1996); H. L. Brown, E. Dan Dahlberg, M. Kief, and Ch. Hou, J. Appl. Phys. **91**, 7415 (2002)
13. X. Martí, F. Sánchez, D. Hrabovsky, J. Fontcuberta, V. Laukhin, V. Skumryev, M.V. García-Cuenca, C. Ferrater, M. Varela, U. Lüders, S. Estradé, J. Arbiol, and F. Peiró, J. Appl. Phys, submitted
14. M. Fiebig, D. Frölich, Th. Lottermoser, and R.V. Pisarev, Phys. Rev. B **65**, 224421 (2002)

**Figure captions**

**Figure 1.** (**a**) Magnetization loops of Py/YMnO$_3$/Pt, measured at 2 K, after cooling the sample from room-temperature in 3 kOe field, under various bias-voltage (V$_e$) values. The circle and arrow illustrate schematically the expected change of magnetization when biasing the sample by an electric field. Inset: Temperature dependence of the magnetization at H = -100 Oe and V$_e$ = 0, when heating the sample from 2 K to 25 K (top panel) and subsequent cooling-heating-cooling cycles from 25 K to 2 K (bottom panel). (**b**) Dependence of the magnetization on the bias-voltage (V$_e$) measured at 2 K in -100 Oe field after cooling the sample from room-temperature in 3 kOe field. Insets: left - zoom of the +1.2 V to -1.2 V and -1.2 V to 0 portions of the bias excursion; right - the sketch of the sample structure and electric biasing.

**Figure 2.** Angular dependence of the magnetoresistance of the Py film at different temperatures. Measurements were done using a magnetic field of 40 Oe, after field-cooling (3 kOe) applied at 45º with respect to the measuring current direction.

**Figure 3.** Angular dependence of the magnetoresistance of the Py film when biasing the Py/YMnO$_3$/Pt sandwich by an electric field. Measurements were done at 5 K, after 3 kOe field-cooling, applied at 45º with respect to the measuring current, from room temperature and applying V$_e$ = 0, 1.2 and 1.8 V bias voltages.

**Figure 4.** Two dimensional map of the angular dependence of the magnetoresistance as a function of the bias-voltage, from +1.8 V to -1.8 V. Scale bar is in Ω.

**Figure 5.** Scheme of the uniaxial in-plane magnetic anisotropy for zero electric bias field (a) and a larger one (b). The orientation of the **M** with respect to the measuring field (**H$_a$**), the field cooling direction (**H$_{eb}$**), and the current (**J**) is displayed. Bottom panels show the calculated AMR response in each case according to Eq. (1). See the text for further details.

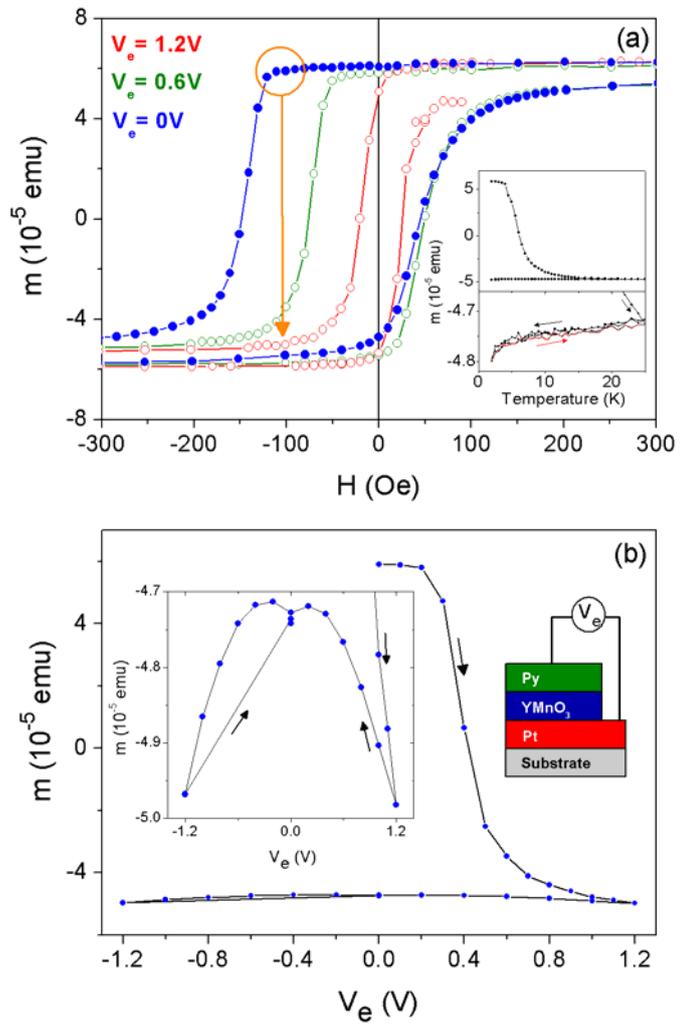

**Figure 1**

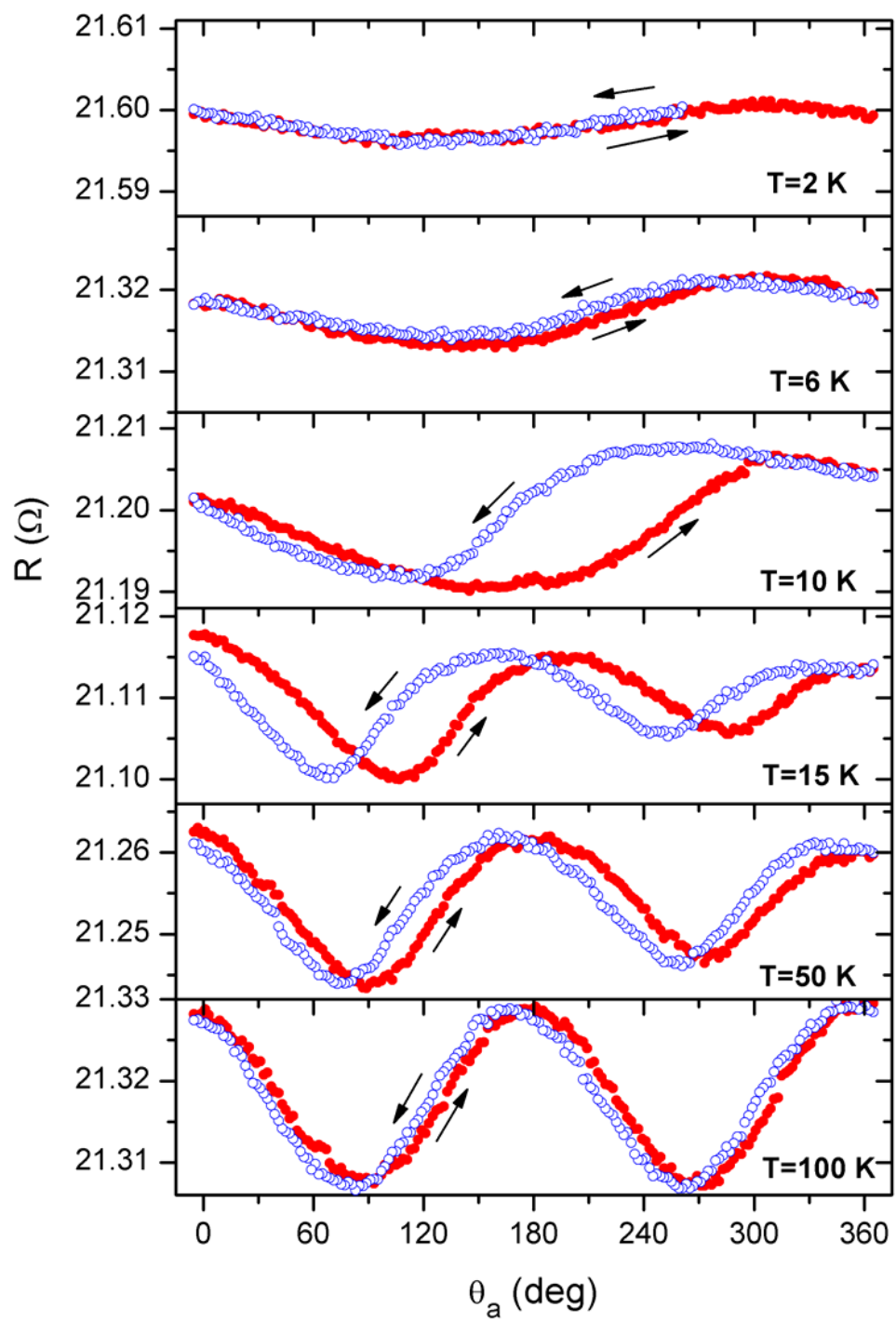

**Figure 2**

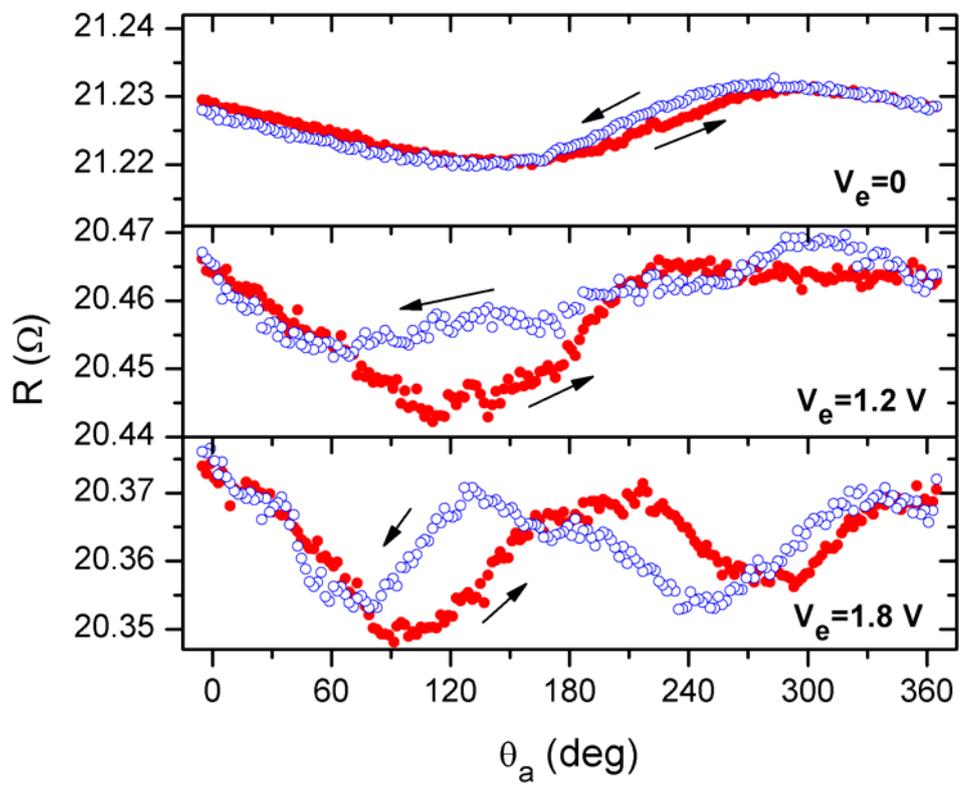

**Figure 3**

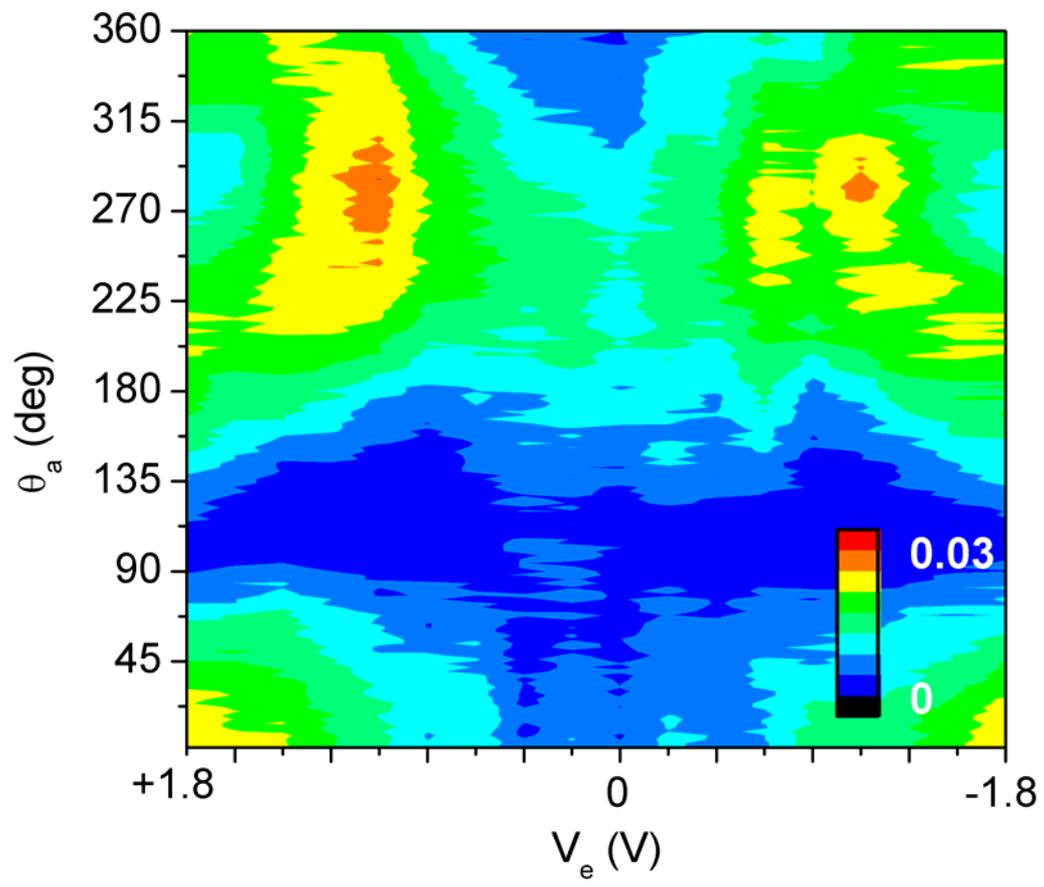

**Figure 4**

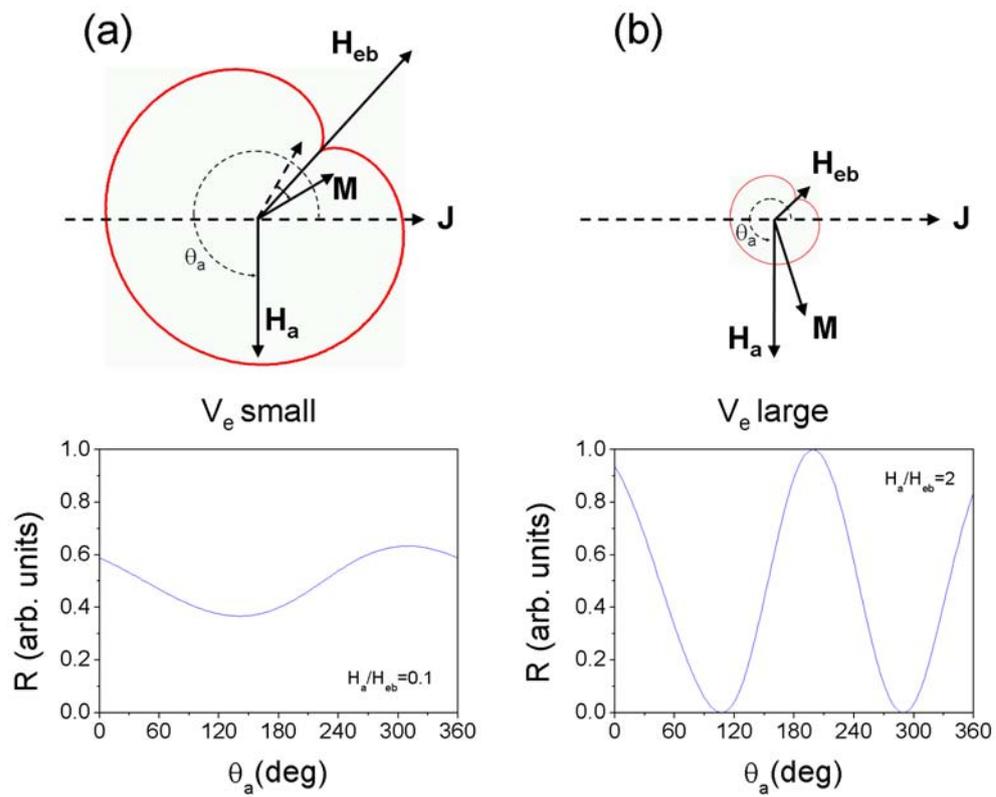

**Figure 5**